\documentclass{aa}
\usepackage{graphicx}
\usepackage{txfonts,color}
\begin{document}
   \title{Large-scale horizontal flows in the solar photosphere I}

   \subtitle{Method and tests on synthetic data}

   \author{M. \v{S}vanda
          \inst{1,2}
          \and
          M. Klva\v{n}a
          \inst{1}
          \and
          M. Sobotka\inst{1}
          }

   \offprints{M. \v{S}vanda}

   \institute{Astronomical Institute, Academy of Sciences of the Czech Republic, Fri\v{c}ova 298,
              CZ-251 65, Ond\v{r}ejov, Czech Republic\\
              \email{svanda@asu.cas.cz, mklvana@asu.cas.cz, msobotka@asu.cas.cz}
         \and
             Astronomical Institute, Charles University in Prague, V Hole\v{s}ovi\v{c}k\'{a}ch 2,
             CZ-180 00, Prague~8, Czech Republic\\
             }

   \date{Received ; accepted }
  \abstract{We propose a useful method for mapping large-scale velocity fields in the solar photosphere. It is based on the local
   correlation tracking algorithm when tracing supergranules in full-disc dopplergrams. The method was developed using synthetic data.
   The data processing the data are transformed during the data processing into a suitable coordinate system, the noise is removed, and finally the velocity field is
   calculated. Resulting velocities are compared with the model velocities and the calibration is done. From our results it becomes clear that
   this method could be applied to full-disc dopplergrams acquired by the Michelson Doppler Imager (MDI) onboard the Solar and Heliospheric Observatory (SoHO).
 
   }
\titlerunning{Large-scale horizontal flows in the solar photosphere I: Method and tests on synthetic data}

   \maketitle
%
\section{Introduction}

The solar photosphere is a very dynamic layer of the solar atmosphere. It is strongly influenced by the underlying convection zone. Despite years of intensive studies, the velocity fields in the solar photosphere remain not very well known. The evidence of the vigorous and sometimes chaotic character of the motions of observed structures in the photosphere (sunspots, granules, and other features) came already from the first systematic studies made in 19th century, of which let us at least mention the discovery of the solar differential rotation (Carrington \cite{carrington}). Motions in the photosphere are strongly coupled with magnetic fields. The large-scale velocity fields are very important for studies of the global solar dynamo. 

An attempt to describe the differential rotation by a parabolic dependence did not make for clear results. Coefficients of the parabola differed according to traced objects and also changed with time, when tracing one type of object (reviewed e.~g. by Schr\"oter \cite{schroeter}). The character of the differential rotation has never been in doubt.

It follows from these arguments that a temporally variable streaming of the plasma exists on the surface of the Sun, which can be roughly described by the differential rotation. This streaming has a large-scale character, large-scale plasma motions were studied for example on the basis of tracking the magnetic structures (e. g. Ambro\v{z} \cite{ambroz_a}, \cite{ambroz_b}). The long-term Doppler measurements done by the MDI onboard SoHO make it possible to extend the studies of large-scale velocities in the solar photosphere. The knowledge of the behaviour of velocities in various periods of the solar activity cycle could contribute to understanding the coupling between the velocity and magnetic fields and of the solar dynamo function.

There are at least three methods calculating photospheric velocities:

\begin{figure*}
\resizebox{0.5\textwidth}{!}{\includegraphics{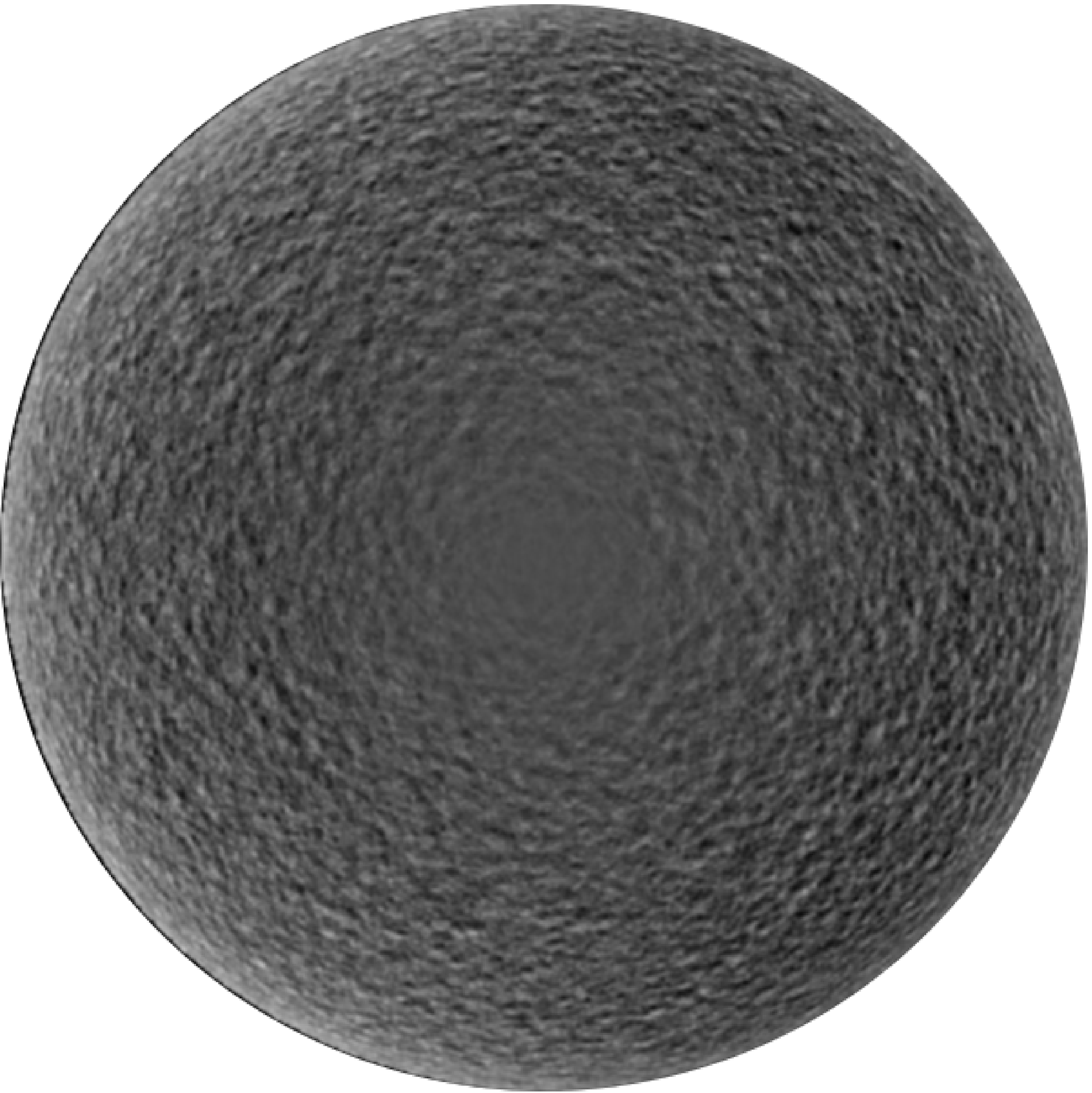}}
\resizebox{0.5\textwidth}{!}{\includegraphics{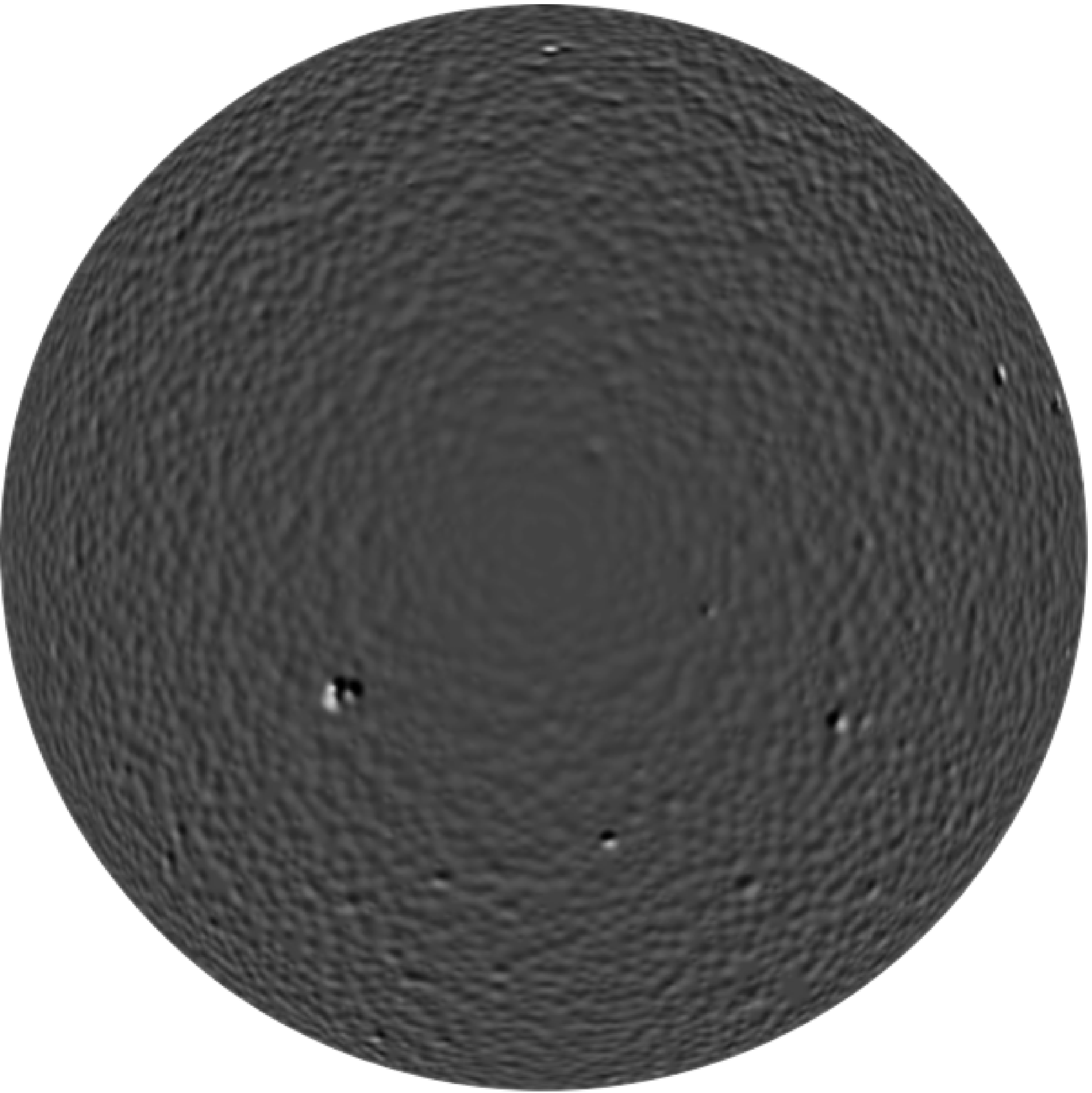}}
\caption{Comparison of the real dopplergram (left, observed by MDI/SoHO) and
the simulated one (right). Both images are visually very similar. The black colour means line-of-sight velocity $-700$~m\,s$^{-1}$ (towards an observer) while white represents $+700$~m\,s$^{-1}$.}
\label{svanda_fig:dopplergrams}
\end{figure*}

\begin{enumerate}
\item \emph{Direct Doppler measurement} -- provides only one component (line-of-sight) of the velocity vector. These velocities are generated by local photospheric structures, amplitudes of which are significantly greater than amplitudes of the large-scale velocities. The complex topology of such structures complicates an utilisation for our purpose. Analysing this component in different parts of the solar disc led to very important discoveries (e. g. supergranulation -- Hart \cite{hart} and Leighton et al. \cite{leighton62}).

\item \emph{Tracer-type measurement} -- provides two components of the velocity vector. When tracing some photospheric tracers, we can compute the local horizontal velocity vectors in the solar photosphere. Tracking motions of sunspots across the solar disc led to the discovery of the differential rotation (Carrington \cite{carrington}).

\item \emph{Local helioseismology} -- provides a full velocity vector. Although local helioseismology (e. g. Zhao \cite{zhao}) is a very promising method, it is still in progress and until now does not provide enough reliable results.
\end{enumerate}

Since the photosphere is a very thin layer (0.04~\% of the solar radius), the large-scale photospheric velocity fields have to be almost horizontal. Then, the tracer-type measurement should be sufficient for mapping the behaviour of such velocities. In this field the local correlation tracking (LCT) method is very useful.

This method was originally designed for the removal of the seeing-induced distortions in image sequences (November \cite{november86}) and later used for mapping the motions of granules in the series of white-light images (November \& Simon \cite{november88}). The method works on the principle of the best match of two frames that record the tracked structures at two different instants. For each pixel in the first frame, a small correlation window is chosen and is compared with a somewhat displaced window of the same size in the second frame. A vector of displacement is then defined as a difference in the coordinates of the centres of both windows when the best match is found. The velocity vector is calculated from this displacement and the time lag between two frames.

LCT was recently used for tracking many features in various types of observations, especially for tracking the granules in high-resolution white-light images (e. g. Sobotka et al. \cite{sobotka99}, \cite{sobotka00}). 

The method needs a tracer -- a significant structure recorded in both frames, the lifetime of which is much longer than the time lag between the correlated frames. We decided to use the supergranulation pattern in the full-disc dopplergrams, acquired by the MDI onboard SoHO. We assume that supergranules are carried as objects by the large-scale velocity field. This velocity field is probably located beneath the photosphere, so that the resulting velocities will describe the dynamics in both the photospheric and subphotospheric layer. The existence of the supergranulation on almost the whole solar disc (in contrast to magnetic structures) and its large temporal stability make the supergranulation an excellent tracer. 

The resulting velocities cover the whole solar disc. Hence the velocity field also describes a motion of the supergranulation in the areas occupied by active regions or by magnetic field concentrations. This fact allows the results to be used to study the mutual motions of substructures like sunspots, magnetic field in active regions, background fields, and the quiet photosphere.

Supergranules are structures with a strong convection coupling. The mean size of the supergranular cell is approx. 30~Mm (e. g. Wang \& Zirin \cite{wang}), with the size and shape dependent on the phase of the solar cycle (e. g. S\' ykora \cite{sykora}). Supergranules are quite stable with a mean lifetime of approx. 20~hours (e. g. Leighton \cite{leighton64}). Information about the distribution function of the lifetime was published by  DeRosa et al. (\cite{derosa}). The internal velocity field in the supergranular cell is predominantly horizontal (e. g. Hathaway et al. \cite{hathaway}) with the amplitude approx. 300~m\,s$^{-1}$. Due to the horizontality of the internal velocity field, the supergranules can be observed in dopplergrams on the whole solar disc except for its centre.

We do not propose that this method capable of measuring velocities of order 1~m\,s$^{-1}$, but we do expect that the large-scale velocities will have magnitudes at least one order greater. We also have to take the largest-scale velocities into account like the differential rotation or meridional circulation, which have velocities of at least 10~m\,s$^{-1}$. If for example we take differential rotation described by the formula $\omega {\rm[deg/day]}= 13.064-1.69 \sin^2 b-2.35 \sin^4 b$ (Snodgrass \cite{snodgrass}), we have to expect a velocity approx. 190~m\,s$^{-1}$ in $b=60\,^\circ$ in the Carrington's coordinate system. The main goal of our study (and the proposed method of mapping the velocities should be proxy for this purpose) is to separate the superposed velocity field into the components and to investigate their physics.

This paper is the first one of a series about the large-scale photospheric velocity fields. In the following papers we shall apply the method here suggested to observed data and describe the properties of the large-scale velocity fields in different periods of the solar activity cycle.

\section{Synthetic data}
\begin{figure*}
\resizebox{0.50\textwidth}{!}{\includegraphics{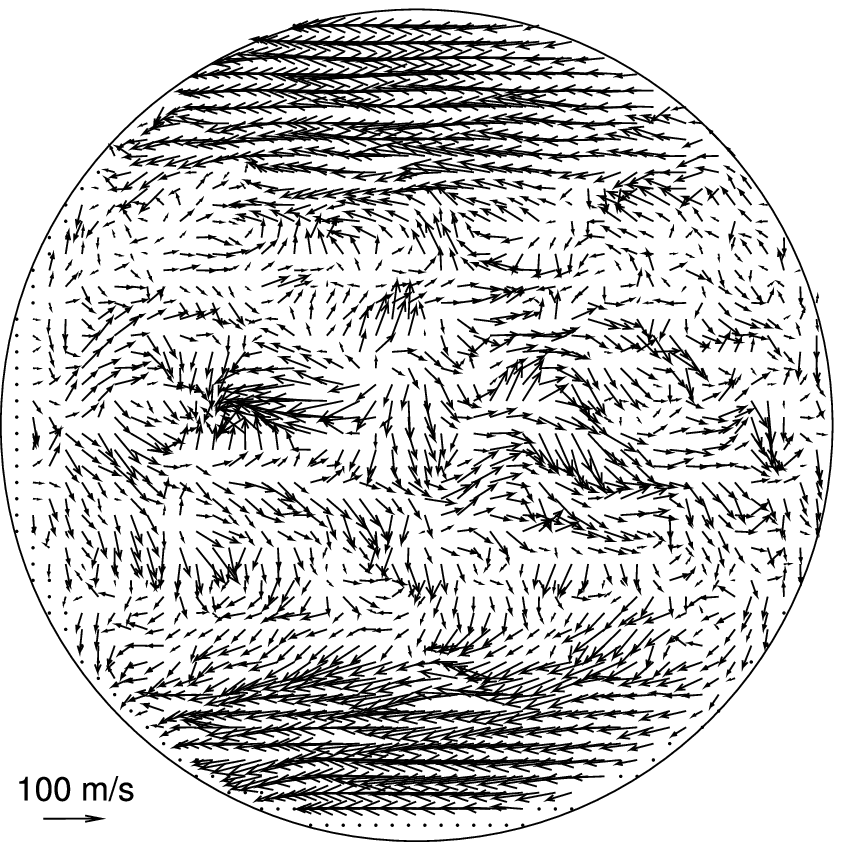}}
\resizebox{0.50\textwidth}{!}{\includegraphics{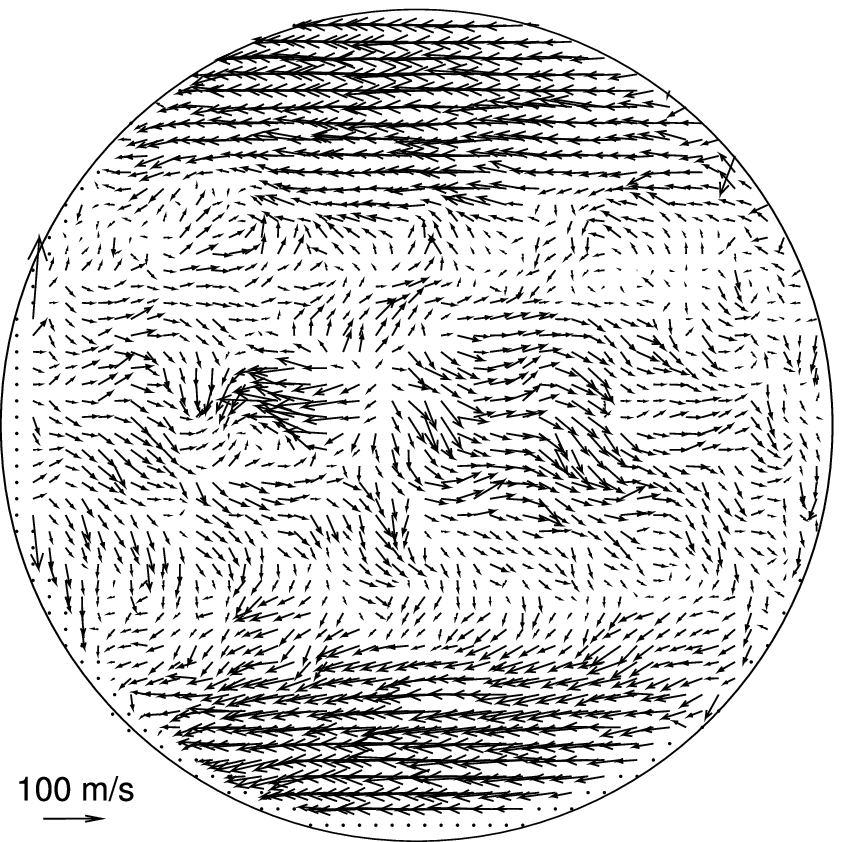}}
\caption{Left -- model vector velocity field. Right -- a velocity field that was computed by applying the LCT method to the synthetic supergranulation pattern with the imposed model field. The arrow lengths, representing the velocity magnitudes, have the same scale. The images are visually very similar, however the magnitudes of calculated velocities are underestimated.}
\label{svanda_fig:sipky}
\end{figure*}

A recent experience with applying this method to observed data (e. g. \v{S}vanda et al. \cite{svanda}) has shown that for the proper setting of the parameters and for the tuning of the method, synthetic (model) data with known properties are needed. The synthetic data for the analysis come from a simple simulation (SISOID code = \emph{SI}mulated \emph{S}upergranulation as \emph{O}bserved \emph{I}n \emph{D}opplergrams) with the help of which we can reproduce the supergranulation pattern in full-disc dopplergrams.

The SISOID code is not based on physical principles taking place in the origin and evolution of supergranulation, but instead on a reproduction of known parameters that describe the supergranulation. Individual synthetic supergranules are characterised as centrally symmetric features described by their position, lifetime (randomly selected according to the measured distribution function of the supergranular lifetime -- DeRosa et al. \cite{derosa}), maximal diameter (randomly according to its distribution function -- Wang \& Zirin \cite{wang}) and characteristic values of their internal horizontal and vertical velocity components (randomly according to their distribution function -- Hathaway et al. \cite{hathaway}).

The most important simplification in the SISOID code is that individual supergranules do not influence each other, but simply overlap. The final line-of-sight velocity at a certain point is given by the sum of line-of-sight velocities of individual synthetic supergranules at the same position.

New supergranules can arise inside the triangle of neighbouring supergranules (identification of such triangles is done by the Delaunay triangulation algorithmed by Barry \cite{barry}) only when the triangle is not fully covered by other supergranules and when anyone of the supergranules located at the vertices of the triangle is not too young, so that in the future it could fully cover the triangle. The position of the origin of the new supergranule is the centroid of the triangle; each vertex is weighted by the size of its supergranule.

The SISOID simulation is done in the pseudocylindrical Sanson-Flamsteed coordinate grid (Calabretta \& Greisen \cite{calabretta}); the transformation from the heliographic coordinates is given by the formula:
\begin{equation}
x=\vartheta \cos \varphi,\ y=\varphi,
\label{svanda_eq:sanson}
\end{equation}
where $x$ and $y$ are coordinates in the Sanson-Flamsteed coordinate system, and $\vartheta$ and $\varphi$ are heliographic coordinates originating at the centre of the disc. At each step an appropriate part of the simulated supergranular field is transformed into heliographic coordinates. The output of the program is a synthetic dopplergram of the solar hemisphere in the orthographical projection to the disc. We assume in our simulation that the Sun lies in an infinite distance from the observer and that the $P$ (position angle of solar rotation axis) and $b_0$ (heliograhic latitude of centre of solar disc) angles are known.

Each step in calculation includes the evaluation of the parameters of individual supergranules, then small old supergranules under the threshold (2~Mm in size) are removed from the simulation and all the triangles are checked, whether a new supergranule can arise inside them. This step in the SISOID code corresponds to 5~minutes in real solar time. The computation is always started from the regular grid. Properties of ``supergranules'' are chosen randomly according to their real distribution functions. The first 1000 steps are ``dummy'', i.~e., no vector velocity field is included and no synthetic dopplergram is calculated. This starting interval is taken for the stabilisation of the supergranular pattern. In the next steps, the model vector velocity field has already been introduced. This field influences only the positions of individual cells. The dopplergram is calculated every third step. For one day in real solar time, 96 dopplergrams are calculated.

The model velocity field with Carrington rotation added is applied according to the assumption of the velocity analysis that the supergranules are carried by a velocity field on a larger scale. In the simulation only, the position of individual supergranules is influenced, and no other phenomena are taken into account. These synthetic dopplergrams are visually similar to the real observed dopplergrams (see Fig.~\ref{svanda_fig:dopplergrams}).

\section{Method of data processing}

The MDI onboard SoHO acquired the full-disc dopplergrams at a high cadence in certain periods of its operation  -- one observation per minute. These campaigns were originally designed for studying the high-frequency oscillations. The primary data contain lots of disturbing effects that have to be removed before ongoing processing: the rotation line-of-sight profile, $p$-modes of solar oscillations. We detected some instrumental effects connected to the data-tranfer errors. It is also known that the calibration of the MDI dopplergrams is not optimal (e. g. Hathaway et al.~\cite{hathaway}) and has to be corrected to avoid systematic errors. While examining long-term series of MDI dopplergrams, we have met systematic errors connected to the retuning of the interferometer. We should also take those geometrical effects into account (finite observing distance of the Sun, etc.) causing bias in velocity determination. According to Strous~(\cite{strous}) for example, the bias coming out of a perspective is about 2~m\,s$^{-1}$, and it depends on the position on the disc. It has been proven (e.~g.~Liu \& Norton~\cite{liu}) that MDI provides reliable velocity measurements when the magnetic field is lower than 2000~Gauss. The velocity observation by MDI will induce up to 100~\% error if the magnetic field is higher than 3000~Gauss due to the magnetic sensitivity of the used Ni\,I line and the limitations of computational algorithm, which cause crosstalks between measured MDI dopplergrams and magnetograms. The removal of these effects will be described in detail in the next paper, while the synthetic data used in this study do not suffer from these phenomena.

As input to the data processing we take a one-day observation that contains 96 full-disc dopplergrams in 15-minute sampling. Structures in these dopplergrams are shifted with respect to each other by the rotation of the Sun and by the velocity field under study.

First, the shift caused by the rotation has to be removed. For this reason, the whole data series (96 frames) is ``derotated'' using Carrington's rotation rate, so that the heliographic longitude of the central meridian is equal in all frames and also equals the heliographic longitude of the central meridian of the central frame of the series. This data-processing step causes the central disc area (``blind spot'' caused by prevailing horizontal velocity component in supergranules) in the derotated series to move with the Carrington's rate. During the ``derotation'' the seasonal tilt of the rotation axis towards the observer (given by $b_0$ -- heliographic latitude of the centre of the disc) is also removed, so that $b_0=0$ in all frames. 

Then the data series is transformed to the Sanson-Flamsteed coordinate system to remove the geometrical distortions caused by the projection of the sphere to the disc. Parallels in the Sanson-Flamsteed pseudocylindrical coordinate system are equispaced and projected at their true length, which makes it an equal area projection. Formulae of the transformation from heliographic coordinates are given by Eq.~(\ref{svanda_eq:sanson}).

The noise coming from the evolutionary changes in the shape of individual supergranules and the motion of the ``blind spot'' in the data series with the Carrington's rotational rate are suppressed by the $k$-$\omega$ filter in the Fourier domain (Title et al. \cite{title}, Hirzberger et al. \cite{hirzberger}). The cut-off velocity is set to 1\,500~m\,s$^{-1}$ and has been chosen on the basis of empirical experience.

The existence of the differential rotation complicates the tracking of the large-scale velocity field, because the amplitudes and directions of velocities of the processed velocity field have a significant dispersion. We have found that, when the scatter of magnitudes is too large, velocities of several hundred m\,s$^{-1}$ cannot be measured precisely by the LCT algorithm where the displacement limit for correlation was set to detect velocities of several tens of m\,s$^{-1}$. Therefore the final velocities are computed in two steps. The first step provides a rough information about the average zonal flows using the differential rotation curve
\begin{equation}
\omega = A+B\sin^2 b+C\sin^4 b\ ,
\label{svanda_eq:fay}
\end{equation}
and calculating its coefficients.

In the second step this average zonal flow is removed from the data series, so that during the ``derotation'' of the whole series the differential rotation inferred in the first step and expressed by~(\ref{svanda_eq:fay}) is used instead of the Carrington rotation. The scatter of the magnitudes of the motions of supergranules in the data transformed this way is much smaller, and a more sensitive and precise tracking procedure can be used.

The LCT method is used in both steps. In the first step, the checked range of velocity magnitudes is set to 200~m\,s$^{-1}$, but the accuracy of the calculated velocities is roughly 40~m\,s$^{-1}$. In the second step the range is only 100~m\,s$^{-1}$ with much better accuracy. The lag between correlated frames equals in both cases 16 frame intervals (i.~e. 4 hours in solar time), and the correlation window with FWHM 30~pixels equals 60\arcsec{} on the solar disc in the linear scale. In one observational day, 80 pairs of velocity maps are calculated and averaged.

For the calculation we use the adapted program {\tt flowmaker.pro} originally written in IDL by Molowny-Horas \& Yi (\cite{molowny}). The algorithm has a limitation in the range of displacements that are checked for each pixel. The quality of correspondence (in our case the sum of absolute differences of both correlation windows) is computed in nine discrete points, then the biquadratic surface is fitted through these nine points, and an extremum position (Darvann \cite{darvann}) is calculated. The final displacement vector is equivalent to the position of the extremum.

\section{Results}

In our tests we have used lots of variations of simple axisymmetric model flows (with a wide range of values of parameters describing the differential rotation and meridional circulation) with good success in reproducing the models. When comparing the resulting vectors of motions with the model ones, we found a systematic offset in the zonal component equal to $v_{\rm offset,\ zonal}=-15\ \rm m\,s^{-1}$. This constant offset appeared in all the tested model velocity fields and comes from the numerical errors during the  ``derotation'' of the whole time series. For the final testing, we used one of the velocity fields obtained in our previous work (\v{S}vanda et al. \cite{svanda}). This field approximates the velocity distribution that we may expect to observe on the Sun. The model flows have structures with a typical size of 60\arcsec{}, since they were obtained with the correlation window of this size.

\begin{figure}[!t]
\resizebox{0.50\textwidth}{!}{\includegraphics{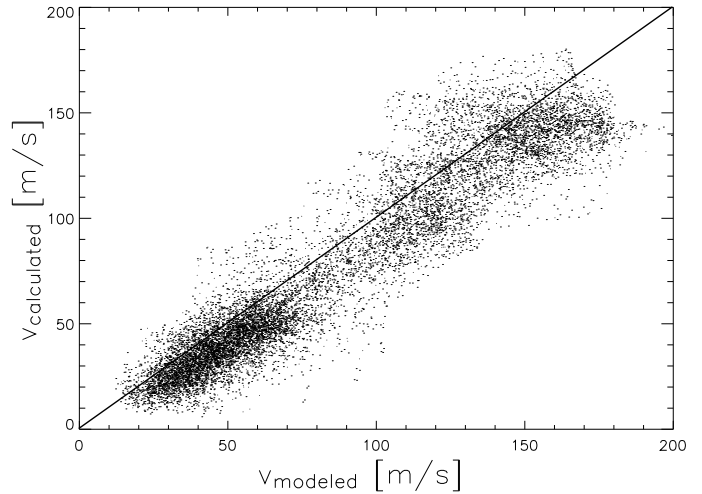}}
\caption{Scatter plot for inference of the calibration curve. Magnitudes of calculated velocities are slightly underestimated by LCT, but the linear behaviour is clearly visible. A line representing the 1:1 ratio is displayed. The calibration affects only the magnitudes of the flows, while the directions do not need any correction.}
\label{svanda_fig:kalibrace}
\end{figure}

\begin{figure}
\resizebox{0.50\textwidth}{!}{\includegraphics{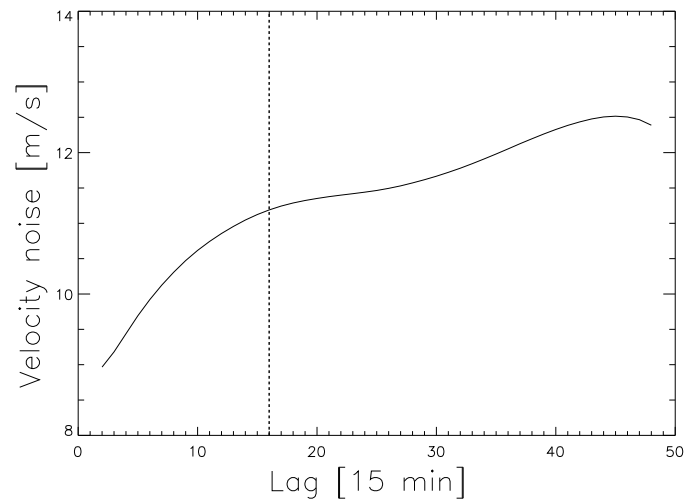}}
\caption{Dependence of the 1-$\sigma$-error of the calculated velocity on the time lag when no model velocity field was introduced and only the evolution of supergranules have been taken into account. The dashed line represents lag 16 (4 hours) that is usually used in our method.}
\label{svanda_fig:noise}
\end{figure}

The calculated velocities (with $v_{\rm offset,\ zonal}=-15\ \rm m\,s^{-1}$ corrected) were compared with the model velocities (Fig.~\ref{svanda_fig:sipky}). Already from the visual impression it becomes clear that most of vectors are reproduced very well in the direction, but the magnitudes of the vectors are not reproduced so well. Moreover, it seems that the magnitudes of vectors are underestimated. This observation is confirmed when plotting the magnitudes of the model vectors versus the magnitudes of the calculated vectors (Fig.~\ref{svanda_fig:kalibrace}). The scatter plot contains more than 1 million points, and most of the points concentrate along a strong linear dependence, which is clearly visible. This dependence can be fitted by a straight line that can be used to derive the calibration curve of the magnitude of calculated velocity vectors. The calibration curve is given by the formula
\begin{equation}
v_{\rm cor}=1.13\,v_{\rm calc},
\label{svanda_eq:calibration}
\end{equation}
where $v_{\rm calc}$ is the magnitude of velocities coming from the LCT, and $v_{\rm cor}$ the corrected magnitude. The directions of the vectors before and after the correction are the same. The uncertainty of the fit can be described by 1-$\sigma$-error 15~m\,s$^{-1}$ for the velocity magnitudes under 100~m\,s$^{-1}$ and 25~m\,s$^{-1}$ for velocity magnitudes greater than 100~m\,s$^{-1}$. The uncertainties of approx. $15\ \rm m\,s^{-1}$ have their main origin in the evolution of supergranules. We studied the dependence of the error of velocity determination on the time lag used when no model velocity field was introduced. We found that this dependence is slowly increasing with the time lag (Fig.~\ref{svanda_fig:noise}) due to the evolution of individual supergranules. Evolution of supergranules is only one part of story, but it gives the lower limit of accuracy that can be obtained by this method. We also ran a test of the LCT sensitivity on the evolution of supergranules when a known underlying velocity field is introduced and came to similar results.

\begin{figure*}[!t]
\resizebox{0.50\textwidth}{!}{\includegraphics{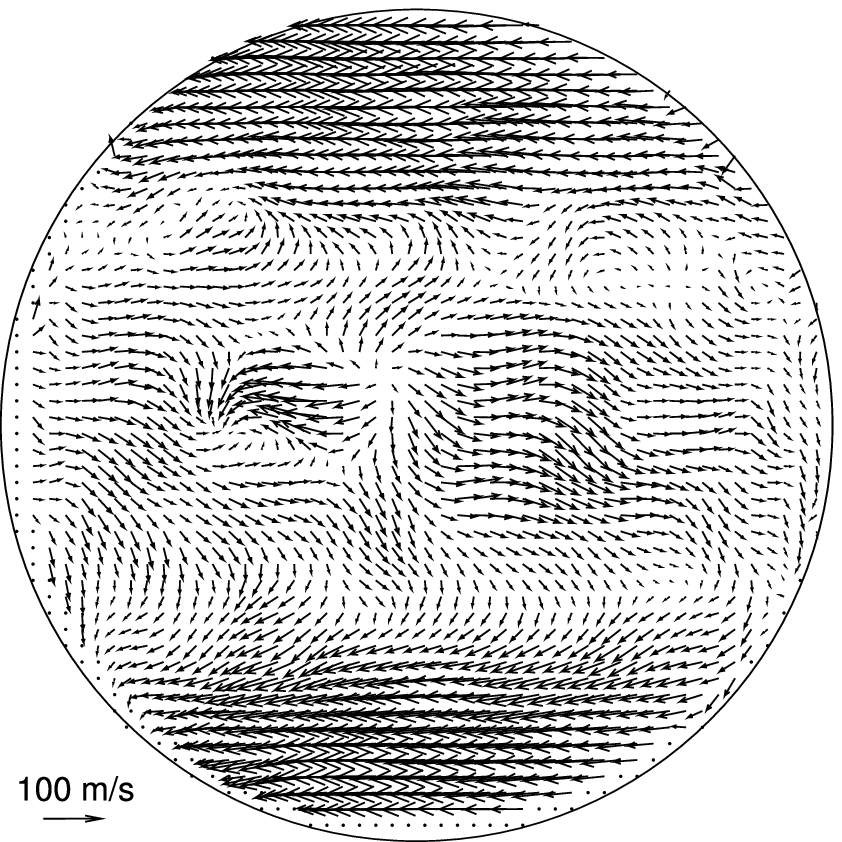}}
\resizebox{0.50\textwidth}{!}{\includegraphics{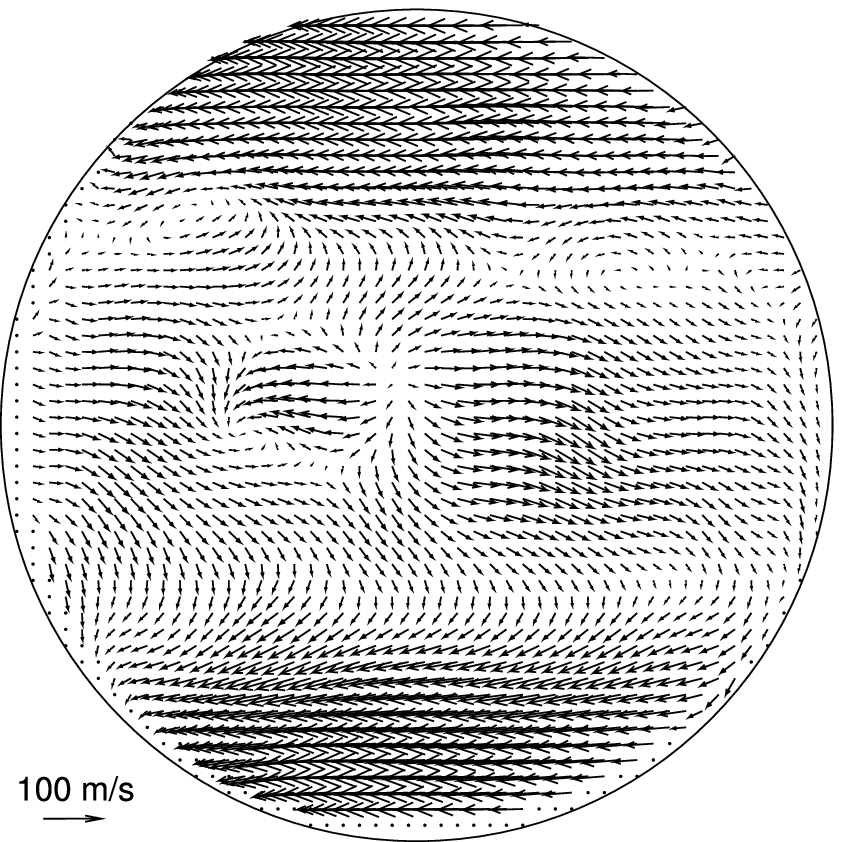}}
\caption{Influence of the calculated velocity field on the choice of FWHM of the correlation window. Left -- 120\arcsec, right -- 200\arcsec. The model velocity field is the same as in Fig.~\ref{svanda_fig:sipky}.}
\label{svanda_fig:fwhm}
\end{figure*}

We tested the sensitivity of the method to the choice of values of FWHM of the correlation window and of the lag between correlated frames in the LCT method. We found that our method is practically insensitive to the choice of the time lag between correlated frames when the lag is in the interval of 10--24 (2.5--6 hours). The larger the lag we choose, the lower the velocities we are able to detect. On the other hand, we have to take into account that a larger lag between correlated frames causes more noise in calculated results coming from evolutionary changes of supergranules and probably also from evolutionary changes in the velocity field under study. According to our tests, for a time lag greater than 30 (7.5~hours), the numerical noise raises very fast. The lag 16 (4 hours) seems to be a good tradeoff between sensitivity and noise. 

\begin{figure}[!b]
\resizebox{0.50\textwidth}{!}{\includegraphics{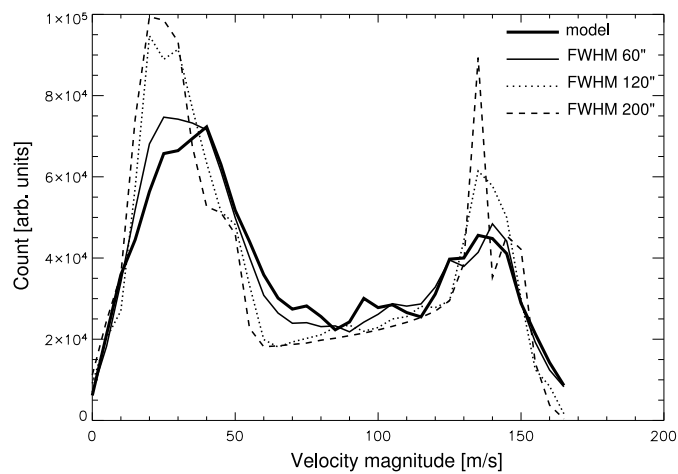}}
\caption{Histograms of velocity magnitudes for various FWHM of the LCT algorithm.}
\label{svanda_fig:histograms}
\end{figure}

The choice of different FWHMs of the correlation window changes the spatial resolution according to FWHM. The general character of the vector field is preserved within the limits of resolution (cf.~Fig.~\ref{svanda_fig:fwhm}). Larger FWHM causes a smoothing of results and an underestimation of vector magnitudes (cf. Fig.~\ref{svanda_fig:histograms}). We found the used parameters (FWHM 60\arcsec, lag 4~hours) to be the best compromise, however these values can be changed during the work on real data.

\section{Conclusions}

We have developed a method for mapping the velocity fields in the solar photosphere based on the local correlation tracking (LCT) algorithm. The method consists in two main steps. In the first main step the mean zonal velocities are calculated and, on the basis of expansion to the Fay's formula (\ref{svanda_eq:fay}), the differential rotation is removed. In the second main step, the LCT algorithm with an enhanced sensitivity is applied. Finally, the differential rotation (obtained in the first step) is added to the vector velocity field obtained in the second main step. Both main steps can be divided into a few substeps, which are mostly common.
\begin{enumerate}
\item ``Derotation'' of the data series using the Carrington's rotation rate in the first step and using the calculated differential rotation in the second step.

\item Transformation of the data series into the Sanson-Flamsteed coordinate system to remove the geometrical distortion caused by the projection to the disc.

\item The $k$-$\omega$ filtering (with the cut-off velocity 1\,500~m\,s$^{-1}$) for suppression of the noise coming from the evolutionary changes of supergranules, of the numerical noise, and for the partial removal of the ``blind spot'' (an effect at the centre of the disc).

\item Application of the LCT: the lag between correlated frames is 4~hours, the correlation window with FWHM 60\arcsec, the measure of correlation is the sum of absolute differences and the nine-point method for calculation of the subpixel value of displacement is used. The calculated velocity field is averaged over the period of one day.
\end{enumerate}

\noindent Finally, the magnitude of the calculated vector field is corrected using formula (\ref{svanda_eq:calibration}) obtained from the tests on the synthetic data.

According to our test, the method provides a very reliable tool for mapping the velocity fields with the spatial resolution of 60\arcsec{} (43,5~Mm) on the solar disc and with the accuracy of 15~m\,s$^{-1}$ for velocity magnitudes under 100~m\,s$^{-1}$ and 25~m\,s$^{-1}$ for velocity magnitudes greater than 100~m\,s$^{-1}$. The method is ready to use on the real full-disc dopplergrams acquired by the MDI onboard SoHO. Our method shows that the usability of the LCT method is wider than originally expected.

\begin{acknowledgements}
The authors of this paper were supported by the Czech Science Foundation under grants 205/03/H144 (M. \v{S}.) and 205/04/2129 (M. K.), by the Grant Agency of the Academy of Sciences of the Czech Republic under grant IAA 3003404 (M. S.) and by ESA-PECS under grant No. 8030 (M. \v{S}.). The Astronomical Institute is working on the Research project AV0Z10030501 of the Academy of Sciences of the Czech Republic. The MDI data were kindly provided by the SoHO/MDI consortium. SoHO is the project of international cooperation between ESA and NASA.
\end{acknowledgements}


\begin{thebibliography}{}

  \bibitem[2001a]{ambroz_a}
    Ambro\v{z}, P. 2001a, \solphys, 198, 253

  \bibitem[2001b]{ambroz_b}
    Ambro\v{z}, P. 2001b, \solphys, 199, 251
  
  \bibitem[1991]{barry}
    Barry J. 1991, Advances in Engineering Software, 13, 325

  \bibitem[1859]{carrington}
    Carrington, R. C. 1859, \mnras, 19, 81
  
  \bibitem[2002]{calabretta}
    Calabretta, M. R., \& Greissen, E. W. 2002, \aap, 395, 1077
  
  \bibitem[1991]{darvann}
    Darvann, T. A. 1991, Ph.D. Thesis, University of Oslo

  \bibitem[2000]{derosa} 
    DeRosa, M. L., Lisle, J. P., \& Toomre, J. 2000, SPD Meeting at Lake Tahoe, article 1.06
  
  \bibitem[1956]{hart}
    Hart, A. B. 1956, \mnras, 116, 38
  
  \bibitem[2002]{hathaway}
    Hathaway, D. H., Beck, J. G., Han, S., \& Raymond, J. 2002, \solphys, 205, 25

  \bibitem[1997]{hirzberger}
    Hirzberger, J., V\' azquez, M., Bonet, J. A., Hanslmeier, A., \& Sobotka, M. 1997, \apj, 480, 406

  \bibitem[1962]{leighton62}
    Leighton, R. B., Noyes, R. W., \& Simon, G. W. 1962, \apj, 135, 474

  \bibitem[1964]{leighton64}
    Leighton, R. B. 1964, \apj, 140, 1547

  \bibitem[2001]{liu}
    Liu, Y., \& Norton, A. A. 2001, SOI Technical Note 01-144, HEPL, Stanford University, 35 pages

  \bibitem[1994]{molowny}
    Molowny-Horas, R., \& Yi, Z. 1994, Internal Report No. 31, Institute of Theoretical Astrophysics, University of Oslo
  
  \bibitem[1986]{november86}
    November, L. J. 1986, \ao, 25, 392

  \bibitem[1988]{november88}
    November, L. J., \& Simon, G. W. 1988, \apj, 333, 427
    
  \bibitem[1985]{schroeter}
    Schr\"oter, E. H. 1985, \solphys, 100, 141
  
  \bibitem[1984]{snodgrass}
    Snodgrass, H. M. 1984, \solphys, 94, 13

  \bibitem[1999]{sobotka99}
    Sobotka, M., V\' azquez, M., Bonet, J. A., Hanslmeier, A., \& Hirzberger, J. 1999, \apj, 511, 436

  \bibitem[2000]{sobotka00}
    Sobotka, M., V\' azquez, M., Cuberes, M. S., Bonet, J. A., \& Hanslmeier, A. 2000, \apj, 544, 1155

  \bibitem[2000]{strous}
  Strous, L. H. 2000, \solphys, 195, 219
  
  \bibitem[2005]{svanda}
  \v Svanda, M., Klva\v na M., \& Sobotka, M. 2005, Hvar Obs. Bull., 29, 39
  
  
    \bibitem[1970]{sykora}
    S\' ykora, J. 1970, \solphys, 13, 292

\bibitem[1989]{title}
    Title, A. M., Tarbell, T. D., Topka, K. P., Ferguson, S. H., Shine, R. A., et al. 1989, \apj, 336, 475

  \bibitem[1989]{wang} 
    Wang, H., \& Zirin, H. 1989, \solphys, 120, 1
  
  \bibitem[2004]{zhao}
    Zhao, J. 2004, Ph.D. Thesis, Stanford University

\end{thebibliography}
\end{document}